\documentclass{aa}  

\usepackage{graphicx}
\usepackage{txfonts}
\usepackage{soul}
\usepackage{booktabs}
\usepackage{multirow}
\usepackage{xcolor}
\begin{document}

\title{Y\,Gem, a symbiotic star outshone by its asymptotic giant branch primary component}

   \author{M.~A.~Guerrero
          \inst{1}
          \and
          D.~A.~Vasquez-Torres\inst{2}
          \and
          J.~B.~Rodr\'{i}guez-Gonz\'{a}lez\inst{2}
          \and
          J.~A.~Toal\'{a}\inst{2}\thanks{Visiting astronomer at the IAA-CSIC as part of the Centro de Excelencia Severo Ochoa Visiting-Incoming programme.}
          \and
          R.~Ortiz\inst{3}
          }
   \institute{Instituto de Astrof\'{i}sica de Andaluc\'{i}a, IAA-CSIC, Glorieta de la Astronom\'{i}a S/N, Granada 18008, Spain\\
         \email{mar@iaa.es}
         \and
         Instituto de Radioastronom\'{i}a y Astrof\'{i}sica, Universidad Nacional Aut\'{o}noma de M\'{e}xico, Morelia, Michoac\'{a}n, Mexico
         \and
         Escola de Artes, Ci\^encias e Humanidades, USP, Av. Arlindo Bettio 1000, 03828-000 S\~ao Paulo, Brazil 
         }

   \date{2024}

 
\abstract{
A considerable number of asymptotic giant branch (AGB) stars exhibit UV excess and/or X-ray emission that indicates a binary companion. 
AGB stars are so bright that they easily outshine their companions. This almost prevents their identification. 
Y\,Gem has been known for some decades to be an AGB star that is bright in the far-UV and X-rays, but it is unclear whether its companion is a main-sequence star or a white dwarf (WD) in a symbiotic system (SySt).
}{
Our goal is to uncover the true nature of Y\,Gem, which will help us to study the possible misidentified population of SySts.
}{
Multiwavelength IR, optical, UV, and X-ray observations were analyzed to investigate the properties of the stellar components and the accretion process in Y\,Gem. 
In particular, an optical spectrum of Y\,Gem is presented here for the first time, while X-ray data are interpreted by means of reflection models produced by an accretion disk and material in its vicinity.
}{
The optical spectrum exhibits the typical sawtooth-shaped features of molecular absorptions in addition to narrow recombination and forbidden emission lines. 
The emission lines and the analysis of the extinction-corrected UV spectrum suggest a hot component with $T_\mathrm{eff}\approx$60,000~K, $L$=140~L$_{\odot}$, and $R$=0.11~R$_{\odot}$ that very likely is an accreting WD. 
{
The late component is found to be an 1.1 M$_\odot$ AGB star with $T_\mathrm{eff}$=3350~K 
and $R$=240~R$_\odot$.}
}{
Using IR, optical, UV, and X-ray data, we found that Y\,Gem is an S-type SySt whose compact component is accreting at an estimated mass-accretion rate of  $\dot{M}_\mathrm{acc}=2.3\times10^{-7}$~M$_\odot$~yr$^{-1}$. 
At this accretion rate, the accreting WD has reached the stable and steady burning phase in which no recurrent events are expected.
}

   \maketitle
    

\section{Introduction}
\nolinenumbers

Planetary nebulae (PNe), the short-lived descendants of low- and intermediate-mass stars after the asymptotic giant branch (AGB), exhibit a high occurrence of axisymmetric morphologies, with bipolar, highly collimated, and point-symmetric shapes \citep{FP2010}.  
The axisymmetric morphology of PNe would be naturally explained by the influence of a companion star during the late AGB phase \citep{BF2002,JB2017}.  
Searches for companions of central stars of PNe (CSPNe) are difficult, but have resulted in a growing sample of binary CSPNe \citep{Miszalski+2009,DeMarco+2013,Jacoby+2021}.

The search for the stellar companions that shape PNe might also be carried out among their precursor AGB stars.  
These are expected to be UV faint \citep{Sahai+2008} and X-ray quiet as they cannot support a corona \citep{LH1979}.  
Therefore, the detection of UV excess and/or X-ray emission was proposed to be evidence of binarity.  
In particular, \citet{OG2016} and \citet{Sahai2022} proposed that AGB stars with a far-UV counterpart, (the so-called fuvAGB stars), high near-UV excess (i.e., $Q_{\rm NUV} > 20$, which is the observed-to-predicted near-UV flux ratio), or $F_{\rm FUV}/F_{\rm NUV}> 0.06$ were most likely in a binary system. 
Similarly, AGB stars with X-ray counterparts (the so-called X-AGB stars) in excess of a few times $10^{29}$ erg~s$^{-1}$ \citep{SK2003,Sahai2015,Ortiz2021} are most likely members of binary systems. 
So far, about 40 fuvAGB stars \citep{Sahai+2008,OG2016} and 50 X-AGB stars \citep[][and references therein]{Guerrero+2024} are known.

One of the most astounding far-UV and X-ray AGB star is Y\,Gem, a 
semi-regular SRb variable with a visual magnitude between 10.4 and 12.3 within a period of 160 d \citep{Samus+2017}.  
The detection of strong far-UV emission \citep{Sahai+2008} was soon followed by the discovery of variable UV and strong and variable X-ray emission \citep{Sahai2011,Sahai2015,Ortiz2021} that was attributed to accretion onto a companion star or to an accretion disk around it.  
The flickering of its UV continuum on timescales $<20$~s strongly supports the presence of an active accretion disk around a companion star, whereas high-velocity absorption and emission components arise from a fast outflow and infalling material from the giant onto the disk \citep{Sahai+2018}.

The nature of the companion star of Y\,Gem, however, is disputed.  
\citet{Sahai+2018} favored a main-sequence companion based on the amount of its UV excess, the relatively low-outflow velocity, and the lack of narrow-band optical emission lines typical of symbiotic stars (SySts), such as H~{\sc i}, He~{\sc ii}, and [O~{\sc iii}] \citep[see also][]{Sahai2011}.  
The X-ray and UV properties of Y\,Gem were also used to suggest that Y\,Gem is actually an SySt \citep{Yu+2022}, in which a WD in a wide binary orbit accretes material from a cool giant companion, an AGB star in this particular case, in the wind Roche-lobe overflow scenario. 
Furthermore, the strong X-ray variability and the unambiguous presence of the 6.4 keV Fe fluorescent emission line is strong evidence of X-ray photon reflection from material in the vicinity of the accreting WD companion \citep[e.g.,][]{Eze2014,Toala2024}, but this important observational fact was overlooked in previous studies of Y\,Gem.

\begin{table*}
\caption{Details of the INT IDS spectroscopic observations of Y\,Gem. }   
\label{tab:obs}   
\tiny   
\centering  
\begin{tabular}{llcccccr}   
\hline\hline    
Start Time & \multicolumn{1}{c}{Grating} & $\lambda_0$ & Spectral range & Spectral Dispersion & Slit Width & $R$ & \multicolumn{1}{c}{Exposure Time} \\
(UTC)   &         &  (\AA)      &   (\AA)     &  (\AA)     & (arcsec) & & \multicolumn{1}{c}{(s)} \\
\hline      
2024-02-17T21:36 & R1200B & 4050 & 3200 -- 4120 & 0.48 & 1.055 & 4220 & 5$\times$120~~~~~~ \\
2024-02-17T21:56 & R1200B & 4950 & 4130 -- 5640 & 0.48 & 1.055 & 5430 &  5$\times$80~~~~~~ \\
2024-02-16T22:52 & R1200V & 5850 & 5650 -- 6180 & 0.48 & 1.055 & 6960 &  5$\times$50~~~~~~ \\
2024-02-16T22:38 & R1200V & 6750 & 6170 -- 7320 & 0.48 & 1.055 & 8790 &  5$\times$50~~~~~~ \\
2024-02-16T22:24 & R1200V & 7650 & 7320 -- 8290 & 0.48 & 1.055 & 9660 &  5$\times$50~~~~~~ \\
\hline   
\end{tabular}
\tablefoot{
The spectral range corresponds to the effective spectral range when the segments of the spectra severely affected by the optics vignetting are excised. 
$R$ denotes the average spectral resolution, $R = \lambda_0$/$\Delta\lambda$. 
}
\end{table*}

\begin{figure*}
\centering
\includegraphics[width=0.95\linewidth]{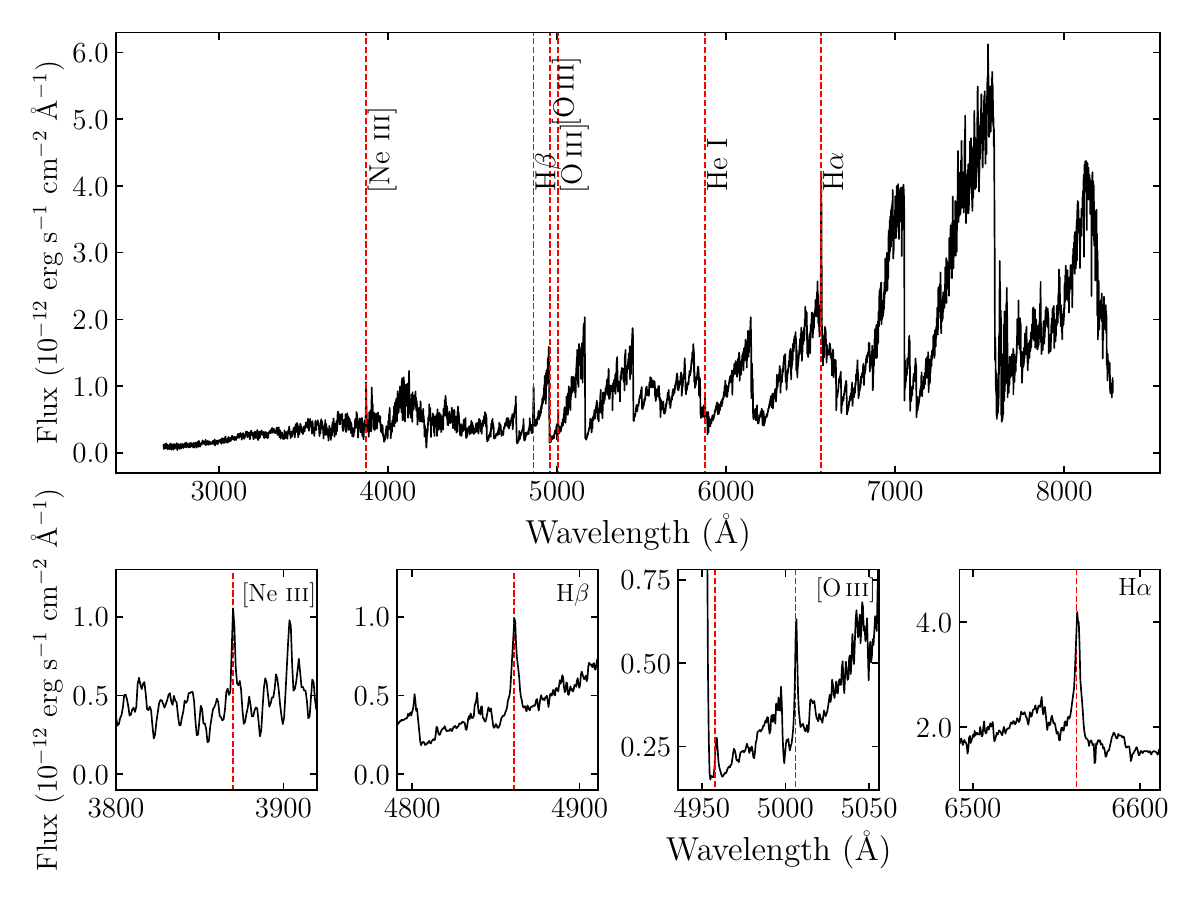}
\caption{
(top) Flux-calibrated INT IDS optical spectrum of Y\,Gem and (bottom) zoomed views of the emission lines of interest.  
}
\label{fig:spectra}%
\end{figure*}

\begin{table*}
\begin{center}
\tiny
\caption{Details of the Chandra ACIS-S and XMM-Newton EPIC-pn observations of Y Gem.}
\setlength{\tabcolsep}{\tabcolsep}  
\begin{tabular}{lcccrrc}
\hline
Instrument  & \multicolumn{1}{c}{Observation ID} & Observation Date  & Epoch  & \multicolumn{1}{c}{Exposure Time}  & \multicolumn{1}{c}{Useful Time} &  Net Count Rate \\
           &         &     (UTC)            & (yr)  & \multicolumn{1}{c}{(ks)}     & \multicolumn{1}{c}{(ks)}        &  (cnt~ks$^{-1}$)   \\
\hline
Chandra ACIS-S & 15714      & 2013-12-15T08:58:59 &  2013.96 & 10.34~~~~~~~     &  10.34~~~~~  &  112$\pm$4 \\ 
Chandra ACIS-S & 16683      & 2014-11-04T07:22:10 &  2014.84 & 9.84~~~~~~~      &   9.84~~~~~  &  90$\pm$3 \\ 
\hline
XMM-Newton EPIC-pn & 0720340201 & 2014-03-31T05:25:26 & 2014.25 &  7.47~~~~~~~ & 2.54~~~~~  & 603$\pm$16 \\
XMM-Newton EPIC-pn & 0763050201 & 2015-09-29T01:52:04 & 2015.74 & 16.74~~~~~~~ & 11.77~~~~~ & 140$\pm$4 \\
XMM-Newton EPIC-pn & 0763050301 & 2015-10-01T01:28:17 & 2015.75 & 16.94~~~~~~~ & 14.93~~~~~ & 100$\pm$3 \\
XMM-Newton EPIC-pn & 0763050401 & 2015-10-23T00:46:36 & 2015.81 & 13.24~~~~~~~ & 9.00~~~~~ & 180$\pm$5 \\
\hline
\end{tabular}
\tablefoot{The net count rates were calculated for the 0.3--10.0 keV energy range.}
\label{tab:xobs}
\end{center}
\end{table*}

The confirmation of Y\,Gem as a SySt \citep[and thus as a binary system with an accreting WD, see][for a recent review on the properties of SySts]{Munari2019} would help to clarify the nature of other AGB stars with far-UV and/or X-ray emission that were suggested to belong to binary systems \citep[e.g.,][]{OGC2019}.  
Despite the statement on the lack of optical forbidden line emission in the optical spectrum of Y\,Gem by \citet{Sahai+2018}, no optical spectra of this star are available.  
Furthermore, the X-ray spectral analysis presented by \citet{Yu+2022} assumed ad hoc Gaussian lines to fit the Fe lines at 6.4, 6.7, and 6.97 keV, but these lines can be used to derive critical information on the physical properties of the accretion onto the disk around the companion star.

This paper presents the first ever optical spectrum of Y\,Gem and an in-depth analysis of publicly available visual photometry (light curve), UV, and X-ray observations to fill these gaps.  
The optical spectrum was used to search for the typical spectral features expected in SySts, including the broad absorption features of C$_2$, CN, VO, and TiO, from the photosphere of the AGB \citep{LanconWood2000}, as well as the narrow emission lines of H~{\sc i}, He~{\sc ii}, and other species in a wide range of ionization stages caused by the strong UV flux of the hot  companion \citep{Munari2019,Akras+2019}.
The analysis of the visual light curve helps us to study the orbital properties of this binary system.
The X-ray spectral properties and variability of Y\,Gem are modeled using consistent reflection physical models developed for X-ray-emitting SySts to infer the properties of the accretion disk, including its inclination to the line of sight and accretion rate.

The photo-geometric distance to Y\,Gem derived by \citet{BailerJones2021} using Gaia DR3 astrometry \citep{Gaia2023} is $d=644^{+51}_{-35}$ pc.  
We note that the values of the goodness-of-fit statistic and RUWE, assessing the quality of the astrometric solution, suggest potential issues for Y\,Gem.  
These may arise from binarity, which is highly probable for Y\,Gem, 
and from variations in the photocenter position caused by the inhomogeneity of the star surface, since the angular diameter of a giant star at this distance would be significantly larger than the parallax itself.  
The latter effects are, however, expected to be small \citep{Beguin+2024}.
Since a previous estimate based on the absolute magnitude of late-M semiregular stars provides a value for its distance of 580 pc \citep{Sahai2011}, only a 10\% lower and within $2\sigma$ of the Gaia distance, the latter can be considered to be reliable and is adopted throughout the paper.

The paper is organized as follows. 
In Section~\ref{sec:obs} we describe the observations we used and their corresponding analysis. 
In Sections~\ref{sec:results} and \ref{sec:diss} we present the results we derived from the analysis of the data and the discussion, respectively. 
Finally, we conclude in Section~\ref{sec:conclusions}.

\section{Observations and data reduction}
\label{sec:obs}

\subsection{Optical spectroscopy}

Optical spectra of Y\,Gem were obtained with the Intermediate-Dispersion Spectrograph (IDS) at the 2.54~m Isaac Newton Telescope (INT) of the Observatorio de El Roque de los Muchachos (ORM; La Palma, Spain) on 2024 February 16 and 17.  
The 235 mm camera and the EEV10 4096$\times$2048 CCD were used, resulting the CCD pixel size of 13.5 $\mu$m on a spatial scale of 0.40 arcsec~pixel$^{-1}$.  
The high-dispersion R1200B and R1200V 
gratings were used to facilitate the detection of narrow emission lines and SySt spectral features against the bright stellar continuum of Y\,Gem.  
The 
gratings were tilted to different central wavelengths to cover the spectral range from 3200 to 8290 \AA.  
Table~\ref{tab:obs} provides details of the different setups, including the central wavelength ($\lambda_0$), the spectral range, the spectral dispersion, the slit width, and the spectral resolution ($R$).

Five exposures of Y\,Gem were obtained for each spectral setup. 
The slit was oriented in all cases along the parallactic angle. 
Observations of the spectro-photometric standard stars GJ\,246 and HD \,19445, 
were obtained immediately after the R1200V and R1200B 
observations of Y\,Gem, respectively.  
The seeing during the observations varied between 2.3 and 3.7$''$ and the sky transparency was photometric.



The spectral optical data were processed using standard {\sc iraf} routines \citep{Tody1993}.  
The individual exposures were combined to remove cosmic rays, and the bias level was subsequently removed.  
The 2D spectra were then flat-field corrected using suitable exposures of a tungsten lamp, and they were corrected for geometrical distortions and were wavelength calibrated using Cu-Ar and Cu-Ne arc lamps. 
One-dimensional spectra were then extracted and flux calibrated using the spectra of GJ\,246 HD \,19445 for the R1200V and R1200B gratings, respectively.  
We note that the IDS optics produce notorious vignetting effects at both extremes of the spectra, which reduces the effective spectral coverage (see Table~\ref{tab:obs} for details). 
The combined spectrum of Y\,Gem in the 3200 to 8200 \AA\ spectral range is presented in Figure~\ref{fig:spectra}.

\subsection{X-ray data}

Y Gem was observed on several occasions with Chandra and XMM-Newton, which allows monitoring its extreme variations in the 0.3--10.0 X-ray band \citep[e.g.,][]{Ortiz2021,Sahai2011,Sahai2015,Yu+2022}. 
The details of the observations used here are presented in Table~\ref{tab:xobs}.  

\begin{figure}
\centering
\includegraphics[width=1.0\linewidth]{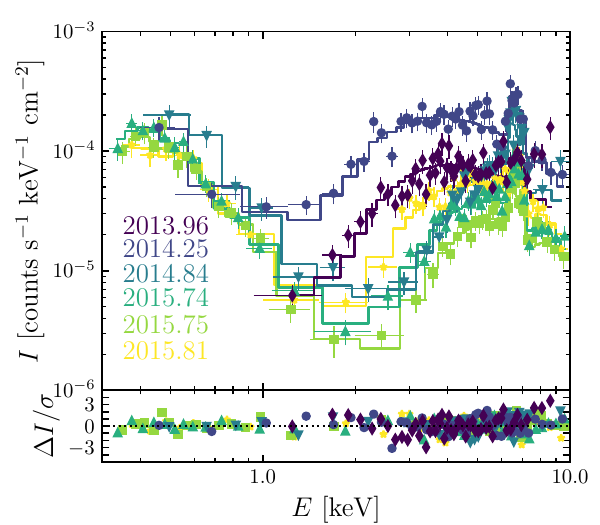}
\caption{Chandra ACIS-S and XMM-Newton EPIC-pn background-subtracted medium-resolution X-ray spectra of Y\,Gem (see Table~\ref{tab:xobs} for details). 
The different symbols represent different epochs. 
The solid lines represent the best-fit models listed in Table~\ref{tab:analysis}, and the fit residuals are shown in the bottom panel.
}
\label{fig:xspec}
\end{figure}

Publicly available Chandra observations of Y Gem (see Tab.~\ref{tab:xobs}) were retrieved from the Chandra Data Archive\footnote{\url{https://cda.harvard.edu/chaser/}}. 
The data were obtained with the Advanced CCD Imaging Spectrometer (ACIS)-S on 2013 December 15 (2013.96) and 2014 November 4 (2014.84) and correspond to Obs. IDs. 15714 and 16683, respectively. These data were processed using standard procedures with the Chandra Interactive Analysis of Observations (CIAO) package \citep{Fruscione2006} version 4.14. 
Spectra were extracted from a circular region with a radius of 8$''$ at the position of Y\,Gem using the CIAO tasks {\it specextract}. 
This task simultaneously created the RMF and ARF calibration matrices.
The background-subtracted ACIS-S spectrum of the 2013.96 and 2014.84 Chandra epochs are presented in Fig.~\ref{fig:xspec}. They were required to have 20 counts per bin at least.

Y~Gem was also observed by XMM-Newton (see Tab.~\ref{tab:xobs}) with the European Photon Imaging Camera (EPIC) in four different epochs: 2014 March 31 (2014.25), 2015 September 29 (2015.74), 2015 October 1 (2015.75), and 2015 October 23 (2015.81). The observation data files were retrieved from the XMM-Newton Science Archive\footnote{\url{https://nxsa.esac.esa.int/nxsa-web/\#search}}. We processed the EPIC data using the Science Analysis
Software \citep[SAS;][]{Gabriel2004} version 20.0 with the most recently updated calibration matrices as of 2024 Jun 28. 
The event files were generated using the {\it epproc} and {\it emproc} SAS tasks. Source spectra were extracted by defining a circular aperture on the position of Y Gem with a radius of 30$''$. The background was extracted from a region in the vicinity without a contribution from any other source. The calibration matrices were produced with the SAS tasks {\it rmfgen} and {\it arfgen}. The background-subtracted EPIC-pn spectra of Y Gem are also presented in Fig.~\ref{fig:xspec}, and they were required to have 30 counts per bin at least. We note that given their superior effective area compared to those of the MOS instruments, only the EPIC-pn data were used for analysis.

\subsection{UV spectra}

For the discussion, we complemented our analysis with publicly available UV spectra of Y\,Gem obtained with the Space Telescope Imaging Spectrograph (STIS) on board the Hubble Space Telescope (HST). The UV spectra were retrieved from the Hubble Legacy Archive\footnote{\url{https://hla.stsci.edu/}}. They correspond to proposal 14713 (PI: R.~Sahai) and were obtained with grisms G140L and G230L, which together cover the $\approx$1120--3200~\AA\, wavelength range. 
The source was observed at two different epochs in October 2016 and April 2017, but its greater brightness on the second epoch prevented a reliable flux calibration of the spectrum \citep[][]{Sahai+2018}.   
After inspecting the spectra, we confirm this issue. 
This limits the analysis to the spectra that were obtained on 2016 October 11-12.
These correspond to total exposure times of 1728.08 s and 1606.81~s for the G140L and G230L grisms, respectively. 
To produce single G140L and G230L spectra, we averaged the available spectra.

\subsection{Optical photometry}

To assess the variability of Y Gem, we retrieved visual magnitude ($\lambda_\mathrm{c}\approx5500$~\AA) photometric data from the American Association of Variable Star Observers\footnote{\url{https://www.aavso.org/}} (AAVSO)
and the $V$ and $g$ magnitudes of the All Sky Automated Survey for SuperNovae\footnote{\url{http://asas-sn.ifa.hawaii.edu/skypatrol}} \citep[ASAS-SN,][]{Shappee2014,Hart2023}. 
The AAVSO data cover the period from epoch 1938.01 to 2024.21, spanning a total of 86.2 years, and the ASAS-SN data cover a much shorter period of $\approx$11 years, from late 2013 to mid 2024. 
Although the AAVSO visual magnitude data do not include uncertainties, they provide the largest baseline of the available photometric measurements, including the ASAS-SN and the AAVSO Photometric All-Sky Survey (APASS).

\begin{figure*}
\centering
\includegraphics[width=\linewidth]{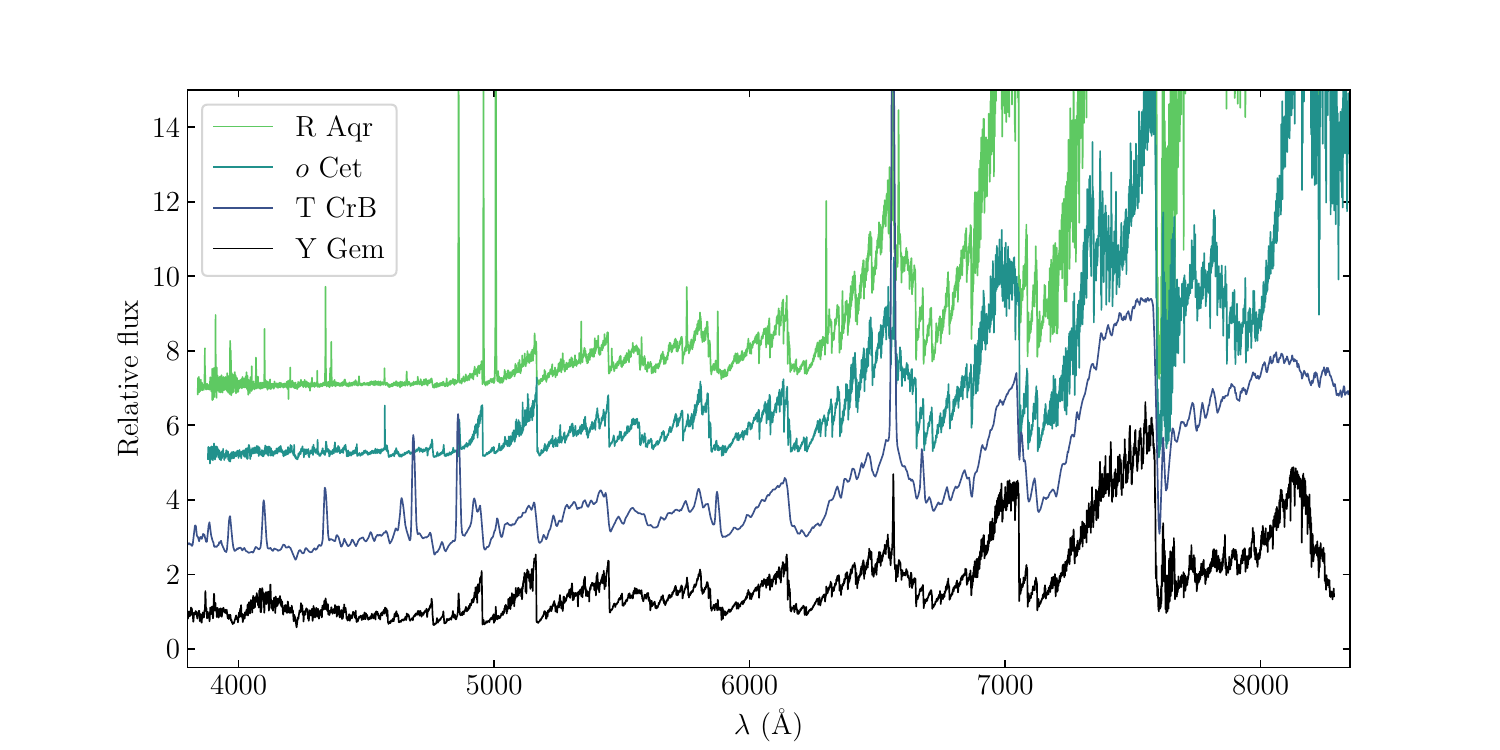}
\caption{Optical spectra of Y\,Gem and the iconic SySts R\,Aqr, $o$\,Cet, and T\,CrB obtained from the ARAS spectral database of eruptive stars. 
The spectra are presented in relative flux, normalized to the intensity of the sawtooth-shaped spectral feature at $\approx$6100 \AA. 
}
\label{fig:comp} %
\end{figure*}

\section{Analysis and results}
\label{sec:results}

\subsection{Optical properties of Y Gem}

The INT IDS optical spectrum of Y\,Gem presented in Fig.~\ref{fig:spectra} shows a number of prominent sawtooth-shaped spectral profiles in the spectral range redward of 4500 \AA.  
These are characteristics of late-type giant stars and are associated with absorption bands of CaH, VO, and TiO, among  others. 
There are also a number of narrow emission lines, which correspond to the H$\beta$, H$\alpha$, [Ne~{\sc iii}] $\lambda$3869 \AA, and [O~{\sc iii}] $\lambda\lambda$4959,5007 \AA\ emission lines (see the bottom panels of Fig.~\ref{fig:spectra}), a tentative detection of He~{\sc i} $\lambda$5876 \AA, and an unidentified narrow emission line at 4777~\AA.  
A high-dispersion ($R \approx 9000$) echelle spectrum of Y\,Gem acquired at Santa Maria de Montmagastrell (F.\ Teyssier, priv.\ comm.) in the framework of the ARAS spectral database of eruptive stars \citep{Teyssier2019}\footnote{See also \url{https://aras-database.github.io/database/index.html}} confirms the presence of the [Ne~{\sc iii}] $\lambda$3869, [O~{\sc iii}] $\lambda$5007, and He~{\sc i} $\lambda$5876 emission lines.  
It adds the detection of [O~{\sc i}] $\lambda\lambda$6300,6363 and [N~{\sc ii}] $\lambda\lambda$5755,6584 narrow emission lines.  
These emission lines, particularly the [Ne~{\sc iii}] and [O~{\sc iii}] emission lines, are indicative of an ionizing source, whereas the nondetection of any other He~{\sc i} emission line besides that of He~{\sc i} $\lambda$5876 is typical of low-ionization SySts.

The optical spectrum of Y\,Gem does not fulfill a number of typical criteria that are used to classify a source as a SySt, such as the presence of
({\it i}) strong He~{\sc ii} $\lambda$4686 \AA\ and emission lines from high-excitation ions (e.g., [Fe~{\sc vii}] $\lambda\lambda$5727,6087 \AA), 
({\it ii}) absorption features TiO $\lambda\lambda$6180,7100 \AA\ and VO $\lambda$7865 \AA, and 
({\it iii}) O~{\sc vi} Raman-scattered lines at 6830 and 7088 \AA\ \citep[e.g.][]{Mikolajewska+1997,Belczynski+2000}.  
Some of these features might be absent in SySts; 19\%, 50\%, and up to 61\% confirmed SySts in the New Online Database of Symbiotic Variables\footnote{https://sirrah.troja.mff.cuni.cz/$\sim$merc/nodsv/} \citep{Merc+2019} lack He~{\sc ii}, O~{\sc vi} Raman-scattered, and [Fe~{\sc vii}] lines, respectively. 
It must be noted in particular that an SySt that hosts a low-temperature WD cannot produce the He~{\sc ii} emission line, whose ionization potential is 54.4 eV. This is far lower than the Raman-scattered O~{\sc vi} lines, which require a higher ionization potential of 113.9 eV. 
Thus, although the detection of Raman-scattered lines at 6830 and 7088 \AA\ is sufficient for a SySt classification \citep{Schmid1989}, it is not necessary. 
Less restrictive classification schemes based on the presence of late-type giant features and strong emission lines of H~{\sc i}, He~{\sc i} and other lines from species with ionization potentials higher than 35 eV (e.g., [O~{\sc iii}]) were proposed by several authors \citep{Belczynski+2000,Miszalski+2013,Merc+2021}. 
Y\,Gem does fulfill these criteria, which might also be the case for a missing population of low-acretion-rate SySts or systems without shell burning that are expected to have weak emission lines \citep[e.g.,][]{Mukai+2016,Munari+2021,Xu+2024}.

\begin{table}
\caption{Flux of the optical emission lines in the spectrum of Y\,Gem.}
\label{tab:lines} 
\tiny
\centering
\begin{tabular}{lc} 
\hline
\hline
Line & \multicolumn{1}{c}{Flux} \\ 
     & \multicolumn{1}{c}{(erg~cm$^{-2}$~s$^{-1}$)} \\
\hline
$[$Ne~{\sc iii}]~3869 & 1.63$\times10^{-12}$ \\ 
? $\lambda$4777 \AA       & 4.17$\times10^{-13}$ \\
H$\beta$~4861         & 2.70$\times10^{-12}$ \\ 
$[$O\,{\sc iii}]~4959 & 2.43$\times10^{-13}$ \\ 
$[$O\,{\sc iii}]~5007 & 7.32$\times10^{-13}$ \\ 
He\,{\sc i}~5876      &                      4.95$\times10^{-13}$:  \\
H$\alpha$~6563        &                      1.02$\times10^{-11}$ \\
\hline
\end{tabular}
\tablefoot{
Fluxes are measured in INT IDS spectroscopic observations obtained with the R1200B and R1200V gratings.  
The flux uncertainty is lower than 10\%, except for the tentative detection of the He~{\sc i} 5876 line.    
}
\end{table}

At any rate, the shape of the spectral continuum of Y\,Gem and the presence of H~{\sc i}, He~{\sc i}, and forbidden emission lines make its optical spectrum similar to that of typical SySts. 
This is illustrated in Fig.~\ref{fig:comp}, which compares the optical spectra of Y\,Gem with those of the iconic SySts R\,Aqr, $o$\,Cet, and T\,CrB obtained from the ARAS spectral database of eruptive stars \citep{Teyssier2019}. 
The similarities in the sawtooth-shaped spectral profiles are quite notorious, while the intensity of the emission lines with respect to the stellar continua varies notably among these different sources and among different observing epochs. 
We particularly note that none of the objects shown in Fig.~\ref{fig:comp} exhibits the Raman-scattered O~{\sc vi} emission lines.

The fluxes of the emission lines detected in the spectrum of Y\,Gem are reported in Table~\ref{tab:lines}. 
The full width at half maximum (FWHM) of the forbidden emission lines is $\approx$2~\AA, which is comparable to the instrumental FWHM.  
In contrast, the FWHM of the Balmer lines is wider, $\approx$4.1~\AA. 
When we adopt recombination Case B, the observed H$\alpha$ to H$\beta$ line ratio $\simeq$3.8 derived from the flux values listed in Table~\ref{tab:lines} implies an extinction value of $E$(B$-$V) = 0.26~mag, that is, $A_\mathrm{V}$=0.80 mag.  
The interstellar reddening toward Y\,Gem, however, is much lower, $E$(B$-$V) $\simeq 0.04$ mag\footnote{
The interstellar extinction vs.\ distance curve towards Y\,Gem provided by Bayestar19 \citep[http://argonaut.skymaps.info,][]{Green+2018} is quite shallow, with values of $E(g-r)$ between 0.02 and 0.06 mag over a wide range of distances around the Gaia distance of 644 pc.  
This results in an average value for $E$(B$-$V) $\simeq 0.04$ mag according to the conversion from $E(g-r)$ to $E(B-V)$ provided by \citet{SF2011}. 
} (or $A_\mathrm{V} \simeq$ 0.12 mag). 
A possible explanation is self-absorption of the Balmer lines, which is very likely in the high-density H~{\sc i}-emitting regions of SySts, particularly in those of S-type stars.  
This results in high values of the intrinsic H$\alpha$ to H$\beta$ line ratio and produces spurious high-extinction values.  
If the observed H$\alpha$ to H$\beta$ line ratio in Y\,Gem were caused by self-absorption, it would imply relatively low values of the H$\alpha$ optical depth, $\tau_{{\rm H}\alpha} \approx 5$ \citep{Netzer1975}.  
To conclude, either the observed H$\alpha$ to H$\beta$ line ratio is affected by mild self-absorption, or the circumstellar material around Y\,Gem reddens the emission that arises from the H~{\sc i}-emitting zone.

\subsection{Light-curve analysis}
\label{lightcurve}

\begin{figure*}
\centering
\includegraphics[width=\linewidth]{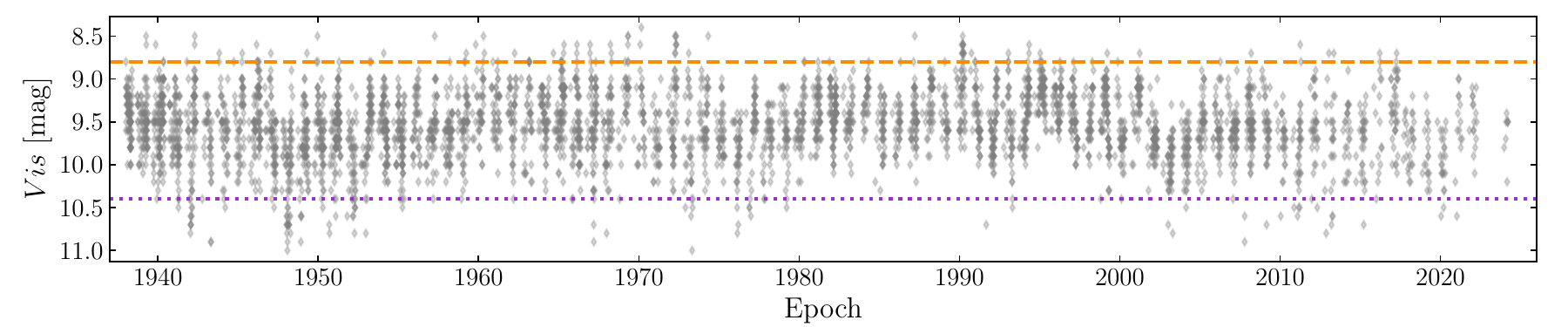}
\caption{
Light curve in the visual band of Y\,Gem obtained from AAVSO. 
The dashed line represents the 2.5th percentile, which defines the minimum magnitude of 8.80 mag, and the dotted line corresponds to the 97.5th percentile and marks the maximum magnitude of 10.40 mag.
The magnitude amplitude between these percentiles is thus 1.6 mag. }
\label{fig:all_dat_aavso}
\end{figure*}

The AAVSO dataset provides the most extended baseline for an identification of both short- and long-term periodic signals that might provide additional evidence of binarity with periods from a few months to decades.
The $\approx$86.2 yr light curve in the visual band of Y\,Gem based on the AAVSO photometric measurements is presented in Fig.~\ref{fig:all_dat_aavso}. 
The minimum and maximum magnitudes, defined by the 2.5th and 97.5th percentiles, are 8.8 mag and 10.4 mag, respectively. 
The light-curve amplitude would be 1.6 mag, which indicates that the cold component of Y\,Gem is not a Mira-type star, but a semiregular (SR) variable \citep[see, e.g.,][]{Samus+2017}.

\begin{figure}
\centering
\includegraphics[width=0.9\linewidth]{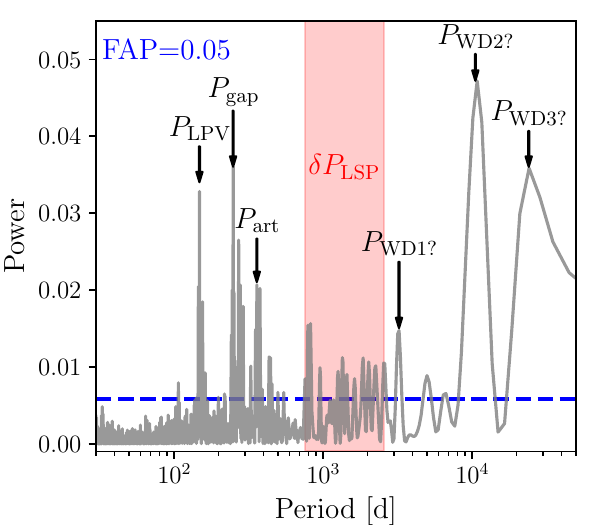}
\caption{Power spectrum of the visual band light curve from AAVSO. The peaks with significant power values are classified as a signal, and the peaks below the blue line, indicating the FAP with a probability of 0.05, are classified as noise.}
\label{fig:ps_LS_aavso}
\end{figure}

Fig.~\ref{fig:ps_LS_aavso} shows the resultant power spectrum (PS)  of the period distribution derived by applying the Lomb-Scargle method \citep{Lomb1976, Scargle1982} to the light curve in Fig.~\ref{fig:all_dat_aavso}.  
Up to six significant discrete periods can be identified in this figure, all of which are above the (blue) dashed line that marks the power threshold probability of 0.05 necessary for a peak (period) to be confidently classified as a signal. The peaks at periods $P_\mathrm{gap} = 250_{-8}^{+40}$ d and $P_\mathrm{art}= 360_{-10}^{+20}$ d are considered spurious.  
The former corresponds to a signal originating from periodic observational gaps, which achieves significant power due to the extensive sampling, whereas the latter is assumed to be an artifact signal due to the Earth's orbital period, which was identified in studies of long-period variable (LPV) stars before \citep[see, e.g.,][]{Fraser2008}.

\begin{figure}
\centering
\includegraphics[width=\linewidth]{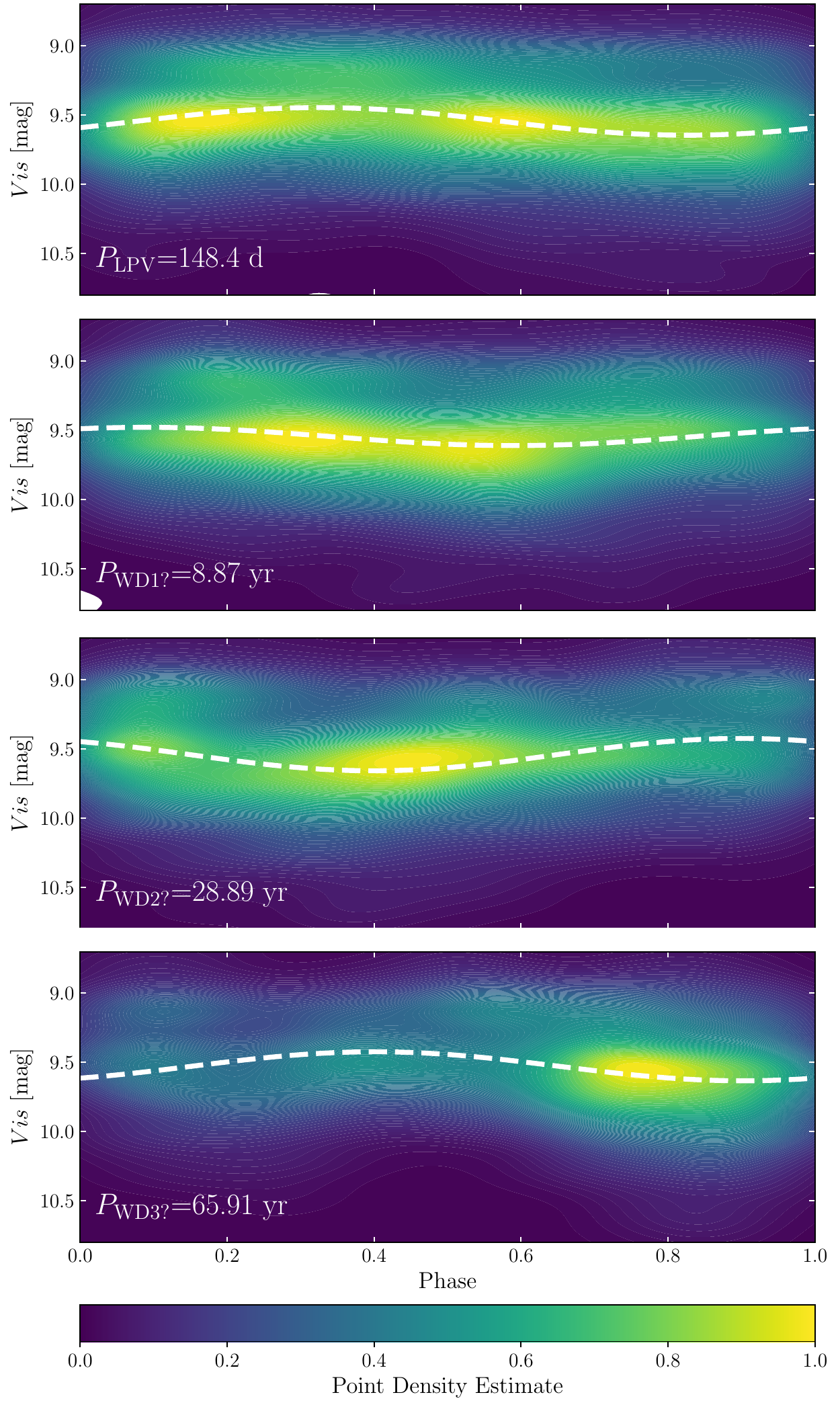}
\caption{
Period-folded light curves of Y\,Gem in the visual band from AAVSO plotted as a function of phase for the most promising periods derived from the LS analysis. 
The dashed white line in each panel represents the corresponding period. 
The density map in each panel has been normalized to fill the range of the displayed model.
}
\label{ps_LS_aavso}
\end{figure}

To confirm that the other four discrete peaks in the PS are real, the period-folded AAVSO light curves are shown in Fig.~\ref{ps_LS_aavso}, phased with the periods of the four additional peaks at 148.4 d and 8.9, 28.9, and 65.9 yr. 
To represent the phased light curves, we used a kernel density estimation \citep[see e.g.,][]{parzen1962} to better visualize the scatter of magnitudes when phased. 
The models derived from the Lomb-Scargle analysis with the identified periods are also displayed. They support the hypothesis that these periods indeed correspond to true signal as the models (dashed white lines) follow regions with a higher point density.

The shortest of these periods at $148.4_{-3.4}^{+9.6}$ d can be interpreted as the pulsation of the LPV M-type star in Y\,Gem.  
This period, which we refer to as $P_\mathrm{LPV}$, is consistent with the value of 160 days reported by \citet{Samus+2017}.
The PS shows then a group of peaks in the 760 to 2570 d range, referred to as $\delta P_\mathrm{LSP}$, which can be associated with the long secondary period (LSP), a phenomenon present in $\approx 25\%$ of LPVs, with values that are five to ten times higher than the primary pulsation mode of the LPV \citep[][]{Olivier2003, wood2004}. 
To date, the mechanism that causes LSPs remains unknown \citep[][]{Pawlak2021}. 
Multiple periods in SR variables are common and normally interpreted as the result of the simultaneous operation of fundamental and overtone pulsation modes \citep{Kiss1999}, but their typical period ratio ($\approx 2.0$) is much shorter than that of the LSPs observed in Y\,Gem.

It is worth mentioning that the $P_\mathrm{LPV}$, $P\mathrm{gap}$, and $P_\mathrm{LSP}$ are recovered in the analysis of the (shorter baseline, $\approx$11 yr) photometric measurements provided by ASAS-SN.  
The combined analysis of the $V$ and $g$ ASAS-SN light curves of Y\,Gem using the multiband Lomb-Scargle periodogram method \citep[][]{VanderPlas2015} of the \texttt{LombScargleMultiband} module of the Python \textit{gatspy} library \footnote{\url{https://www.astroml.org/gatspy/periodic/lomb_scargle_multiband.html}} finds these three peaks with period values of   $P_\mathrm{LPV} = 152.3_{-12.0}^{+3.5}$ d, $P_\mathrm{GAP} = 243_{-10}^{+50}$ d, and $P_\mathrm{LSP} = 1690_{-560}^{+880}$ d. 
The latter thus is in the range from 1130 to 2570 d.

The longest periods in the AAVSO data correspond to a several years and even a few decades.  
These long periods that amount to several years have been observed in at least one-third of the pulsating variables, such as Mira- or SR-type variables.  
They are generally attributed to the existence of a low-mass secondary companion \citep{Soszynski2021} as circumstellar dust matter orbiting the system is asymmetrically distributed in a disk that periodically absorbs a part of the radiation emitted by the primary star and re-emits it in the infrared \citep{WN2009}. 
The shortest of these periods, at $8.87_{-0.40}^{+0.35}$ years, is indeed in the period range from 8.7 to 12.9 yr estimated for the orbit of the WD companion by \citet{Yu+2022}. Thus, we refer to it as $P_\mathrm{WD1?}$.

The periodogram also reveals another two even longer-period peaks denoted as $P_\mathrm{WD2?}$ and $P_\mathrm{WD3?}$. 
The first, $P_\mathrm{WD2?} = 28.9_{-4.2}^{+6.0}$ yr, is the peak with the highest power, and the next $P_\mathrm{WD3?}$ is found at 65.9 yr (with a lower limit of 51.6 yr). 
These can be resonance frequencies of $P_\mathrm{WD1?}$, but we note that SySts such as R\,Aqr or $o$ Cet show somewhat similar long periods of 42.2 \citep[][]{Alcolea2023} and 500 years \citep[][]{Prieur2002}, respectively. 
Alternatively, similarly long periods of around 30 years were attributed to a magnetic activity cycle, like the 11 year cycle of the Sun, but for evolved stars \citep[see e.g., V694\,Mon,][]{Leibowitz2015}. 
This means that the long $P_\mathrm{WD2?}$ and $P_\mathrm{WD3?}$ periods of Y\,Gem might either be associated with the magnetic activity of the giant component of Y\,Gem or with the orbit of its putative WD companion.  
However, these long orbital periods would produce a less efficient mass-accretion process \citep[e.g.,][]{TejedaToala2024}.

\subsection{Reflection physics in Y\,Gem}

As shown by previous authors, Y\,Gem has exhibited dramatic spectral changes in the 0.3--10.0~keV X-ray band \citep[see Fig.~\ref{fig:xspec}; e.g.,][]{Sahai2011,Sahai2015,Ortiz2021,Yu+2022}. 
Y\,Gem initially exhibited a $\delta$-type X-ray spectrum, which, according to the classification scheme defined for X-ray SySts \citep[see][]{Murset1997,Luna2013}, are sources characterized by heavily extinguished hot plasma that mostly emits at energies above 2~keV.
All the other epochs exhibit the clear presence of this extinguished plasma with an additional soft spectral contribution in the 0.3--2.0 keV energy range, which is the signature of $\beta/\delta$-type objects (see Fig.~\ref{fig:xspec}). SySts are known to evolve from one type to the next, in particular, those of $\delta$-type often evolve into $\beta/\delta$-type objects \citep[e.g.,][]{Lucy2020,Toala_T_CrB_2024}.

\begin{figure*}
\centering
\includegraphics[width=\linewidth]{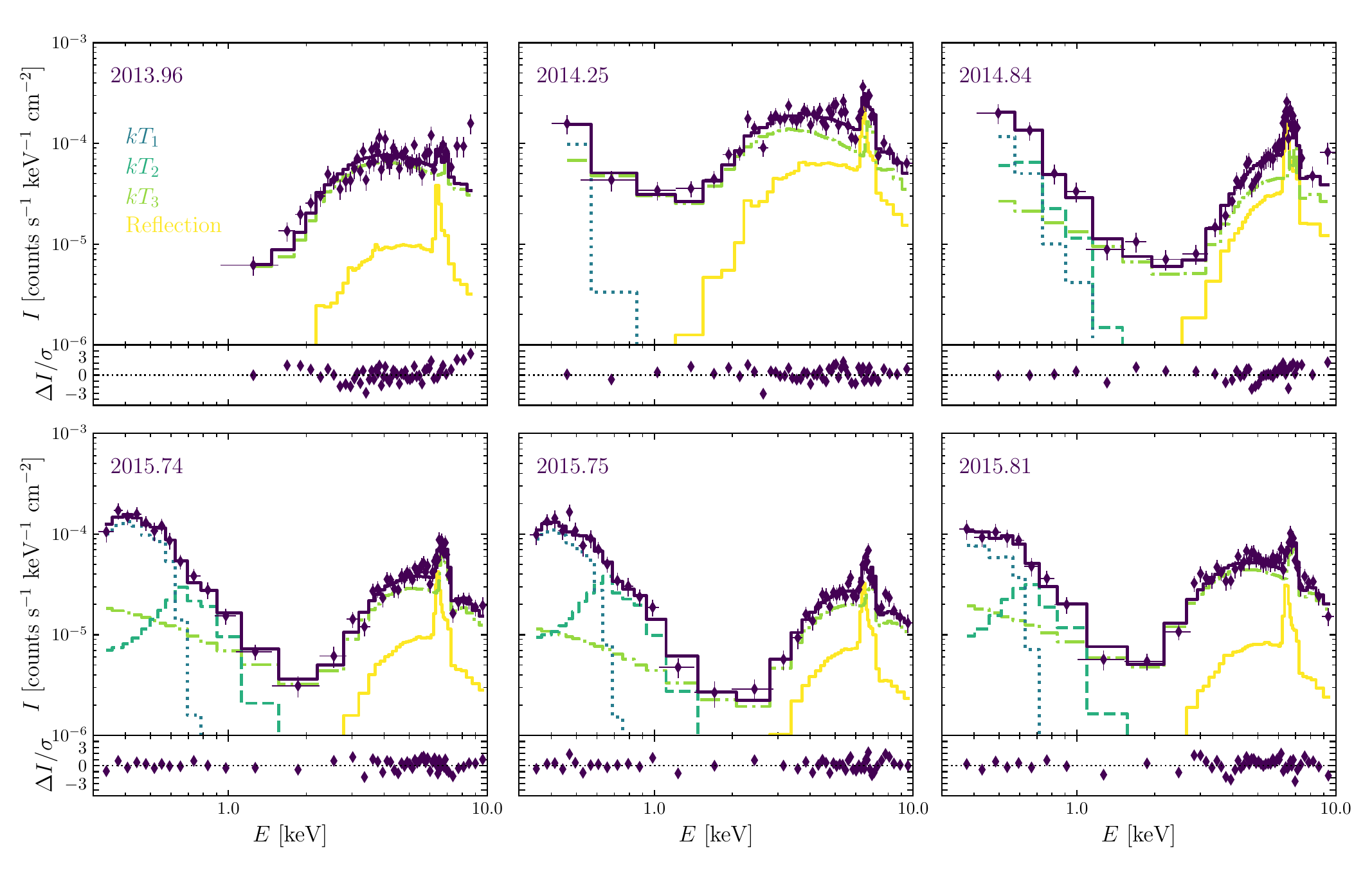}
\caption{Background-subtracted X-ray spectra of Y Gem. The different panels show different epochs with details from their best-fit model spectra (dark solid line). The dotted, dashed, dash-dotted, and light solid lines represent the contributions from the $kT_1$, $kT_2$, $kT_3$, and the reflection components as described in Table~\ref{tab:analysis}.}
\label{fig:xspec2}
\end{figure*}

\begin{table*}
\begin{center}
\tiny
\caption{Best-fit parameters of the spectral analysis of the X-ray observations of Y\,Gem for each available epoch.}
\label{tab:analysis}
\setlength{\tabcolsep}{0.8\tabcolsep}  
\begin{tabular}{lcccccccc}
\hline
\hline
Parameter & Units & 2013.96  & 2014.25 & 2014.84 & 2015.74 & 2015.75 & 2015.81 \\
\hline
$\chi^{2}$/DoF  &    &  83.39/50=1.67 & 54.01/43=1.25 & 170.66/161=1.06 & 170.66/161=1.06 & 170.66/161=1.06 & 170.66/161=1.06\\
\hline
$N_\mathrm{H1}$ & 10$^{22}$ cm$^{-2}$    & \dots & 0.1$\pm$0.08 & 0.42$\pm$0.23 & 0.50$\pm$0.08 & 0.51$\pm$0.07 & 0.41$\pm$0.14\\
$kT_1$          & keV                    & \dots & 0.03$\pm$0.01 & 0.09$\pm$0.04 & 0.04$\pm$0.01 & 0.03$\pm$0.01 & 0.04$\pm$0.01 \\
$A_1$           & cm$^{-5}$              & \dots & 14.1$\pm$0.2 & (5.7$\pm$2.2)$\times10^{-3}$ & 34.6$\pm$2.9 & 72.4$\pm$6.9 & 1.8$\pm$0.3\\
$f_\mathrm{X1}$ & erg~cm$^{-2}$~s$^{-1}$ & \dots & (3.5$\pm$0.4)$\times10^{-13}$ & (3.8$\pm$1.4)$\times10^{-14}$ & (2.8$\pm$0.3)$\times10^{-14}$ & (2.3$\pm$0.2)$\times10^{-14}$ & (1.9$\pm$0.3)$\times10^{-14}$\\
$F_\mathrm{X1}$ & erg~cm$^{-2}$~s$^{-1}$ & \dots & (5.3$\pm$0.1)$\times10^{-12}$ & (1.9$\pm$0.8)$\times10^{-12}$ & (1.6$\pm$0.1)$\times10^{-10}$ & (1.8$\pm$0.2)$\times10^{-10}$ & (2.4$\pm$0.4)$\times10^{-14}$\\
$kT_2$          & keV                    & \dots & \dots & 0.16$\pm$0.01 & 0.21$\pm$0.01 & 0.20$\pm$0.02 & 0.20$\pm$0.01 \\
$A_2$           & cm$^{-5}$              & \dots & \dots & (5.5$\pm$5.2)$\times10^{-4}$ & (1.5$\pm$0.8)$\times10^{-4}$ & (2.4$\pm$1.0)$\times10^{-4}$ & (1.3$\pm$0.8)$\times10^{-4}$\\
$f_\mathrm{X2}$ & erg~cm$^{-2}$~s$^{-1}$ & \dots & \dots & (3.6$\pm$3.4)$\times10^{-14}$ & (1.5$\pm$0.8)$\times10^{-14}$ & (1.8$\pm$0.7)$\times10^{-14}$ & (1.5$\pm$1.3)$\times10^{-14}$\\
$F_\mathrm{X2}$ & erg~cm$^{-2}$~s$^{-1}$ & \dots & \dots & (6.4$\pm$6.3)$\times10^{-13}$ & (3.1$\pm$1.6)$\times10^{-13}$ & (4.7$\pm$1.8)$\times10^{-13}$ & (2.5$\pm$1.5)$\times10^{-13}$\\
\hline
$N_\mathrm{H2}$ & 10$^{22}$ cm$^{-2}$    & 7.5$\pm$0.6   & 5.6$\pm$0.6  & 37.8$\pm$8.2 & 19.7$\pm$1.22 & 24.8$\pm$1.8 & 12.7$\pm$0.8 \\
CF              &                        & {\bf 0.990}   & 0.980$\pm$0.10 & 0.990$\pm$0.003 & 0.990$\pm$0.002 & 0.990$\pm$0.003 & 0.990$\pm$0.002\\
$kT_3$          & keV                    & 29.3$\pm$15.3 & 10.5 $\pm$1.9& 15.3$\pm$4.4  & 10.0$\pm$1.1 & 16.9$\pm$4.0 & 11.3$\pm$1.4\\
$A_3$           & cm$^{-5}$              & (3.4$\pm$0.4)$\times10^{-3}$ &(5.8$\pm$0.9)$\times10^{-3}$ & (5.6$\pm$0.6)$\times10^{-3}$ & (2.7$\pm$0.3)$\times10^{-3}$ & (1.9$\pm$0.2)$\times10^{-3}$ & (3.1$\pm$0.2)$\times10^{-3}$\\
$f_\mathrm{X3}$ & erg~cm$^{-2}$~s$^{-1}$ & (3.3$\pm$0.4)$\times10^{-12}$ & (6.1$\pm$0.9)$\times10^{-12}$ & (2.5$\pm$0.3)$\times10^{-12}$ & (1.7$\pm$0.2)$\times10^{-12}$ & (1.1$\pm$0.1)$\times10^{-12}$ & (2.4$\pm$0.2)$\times10^{-12}$\\
$F_\mathrm{X3}$ & erg~cm$^{-2}$~s$^{-1}$ & (6.8$\pm$0.8)$\times10^{-12}$ & (1.2$\pm$0.2)$\times10^{-11}$ & (1.2$\pm$0.1)$\times10^{-11}$ & (5.7$\pm$0.7)$\times10^{-12}$ & (4.1$\pm$0.3)$\times10^{-12}$ & (6.6$\pm$0.5)$\times10^{-12}$\\
Reflection \\
$A_\mathrm{ref}$& cm$^{-5}$              & 1.10$\pm$0.5 & 6.6$\pm$1.2 & 7.0$\pm$1.1 & 1.5$\pm$0.3 & 1.4$\pm$0.3 & 1.1$\pm$0.3\\
$f_\mathrm{ref}$& erg~cm$^{-2}$~s$^{-1}$ & (5.4$\pm$2.2)$\times10^{-13}$ & (3.5$\pm$0.6)$\times10^{-12}$ & (1.8$\pm$0.3)$\times10^{-12}$ & (5.4$\pm$0.8)$\times10^{-13}$ & (4.1$\pm$1.1)$\times10^{-13}$ & (4.6$\pm$1.2)$\times10^{-13}$\\
$F_\mathrm{ref}$& erg~cm$^{-2}$~s$^{-1}$ & (7.5$\pm$3.7)$\times10^{-12}$ & (4.5$\pm$0.7)$\times10^{-12}$ & (4.7$\pm$0.7)$\times10^{-12}$ & (1.0$\pm$0.2)$\times10^{-12}$ & (9.1$\pm$2.3)$\times10^{-13}$ & (7.5$\pm$1.9)$\times10^{-13}$\\
\hline
$f_\mathrm{X,TOT}$ & erg~cm$^{-2}$~s$^{-1}$& (3.9$\pm$0.6)$\times10^{-12}$ & (9.8$\pm$1.5)$\times10^{-12}$ & (4.4$\pm$0.6)$\times10^{-12}$ & (2.2$\pm$0.4)$\times10^{-12}$ & (1.6$\pm$0.2)$\times10^{-12}$ & (2.9$\pm$0.3)$\times10^{-12}$\\
$F_\mathrm{X,TOT}$ & erg~cm$^{-2}$~s$^{-1}$& (7.5$\pm$1.2)$\times10^{-12}$ & (2.2$\pm$0.2)$\times10^{-11}$ & (1.9$\pm$0.2)$\times10^{-11}$ & (1.6$\pm$0.1)$\times10^{-10}$ & (1.9$\pm$0.1)$\times10^{-10}$ & (3.2$\pm$0.5)$\times10^{-11}$\\
$L_\mathrm{X,TOT}$ & erg~s$^{-1}$   & (3.7$\pm$0.6)$\times10^{32}$ & (1.1$\pm$0.1)$\times10^{33}$ & (9.4$\pm$1.6)$\times10^{32}$ & (7.9$\pm$0.5)$\times10^{33}$ & (9.3$\pm$0.5)$\times10^{33}$ & (1.6$\pm$0.3)$\times10^{33}$\\
\hline
\end{tabular}
\tablefoot{
The observed ($f_\mathrm{X}$) and unabsorbed ($F_\mathrm{X}$) fluxes were computed for the 0.3--10.0~keV energy range. 
The value of CF for the 2013.96 X-ray spectrum in the third column in boldface was fixed.}
\end{center}
\end{table*}

After confirming that the X-ray emission of Y\,Gem is consistent with that of a SySt, we can improve our understanding of the production of X-rays from this binary system. 
We particularly note that using a Gaussian profile to fit the Fe fluorescent line at 6.4 keV is a simplistic observational approximation without a basis in physically motivated models. 
\citet{Toala2023,Toala_T_CrB_2024} recently demonstrated that a considerably fraction of the total observed flux (and luminosity) from the $\beta/\delta$-type SySts CH\,Cyg and T\,CrB originates from jet-like components and reflection at the accretion disk and material in the vicinity of the accreting WD component.
Dissecting the flux from each component provides a tool for improving our estimate of the mass-accretion rate (see Section~\ref{sec:diss}).

The X-ray spectra of Y\,Gem were thus modeled making use of XSPEC, the X-ray Spectral Fitting package  \citep[version 12.12.1;][]{Arnaud1996}. 
Following previous analyses of the X-ray emission from this source \citep[see][and references therein]{Yu+2022}, we required the presence of two soft components with a relatively low hydrogen column density ($N_\mathrm{H}$) in addition to the heavily extinguished plasma component. 
The optically thin plasma emission model {\it apec} included in XSPEC\footnote{\url{http://astroa.physics.metu.edu.tr/MANUALS/xspec12_html/XSmodelApec.html}} was adopted for these spectral components. 
The extinction produced by the hydrogen column density was included by adopting the X-ray absorption model {\it tbabs} \citep{Wilms2000} that is also included in XSPEC. 
The abundances were fixed to solar values \citep{Lodders2009}.

Custom reflection models were then calculated using the stellar kinematics including radiative transfer code SKIRT, which was recently extended to include the
treatment of X-ray photons  \citep[version 9.0;][]{Camps2020}. 
This version of SKIRT includes the effects of Compton scattering on free electrons, photoabsorption, and fluorescence by cold atomic gas, scattering on bound electrons, and extinction by dust \citep{vanderMeulen2023}. 
We adopted flared disk density distributions characterized by an averaged column density $N_\mathrm{H,ref}$, an outer radius $R$, and an inclination angle $\theta$, where $\theta=0^{\circ}$ is the pole-on line of sight of the SySt orbital plane, and $\theta=90^{\circ}$ corresponds to an edge-on view. 
The disk was assumed to have a fixed temperature of 10$^{4}$~K with an inner radius of 8000~km (=5.4$\times10^{-5}$~AU). Different opening angle values between $\varphi$=10 and 30$^{\circ}$ were explored, but they did not produce strong differences. We therefore decided to adopt a fixed value of $\varphi=30^{\circ}$.

Up to 378 SKIRT models were produced covering the $N_\mathrm{H,ref}$=[$10^{23}, 5\times10^{23}, 10^{24}, 5\times10^{24}, 10^{25}, 5\times10^{25}, 10^{26}$]~cm$^{-2}$ and $R$=[0.3, 0.5, 1.0, 1.25, 3.0, 5.0]~AU values. For each disk model, we created  spectral energy distributions (SED) adopting viewing angles between $\theta=10$ and $\theta=90^{\circ}$, with steps of $\Delta\theta=10^{\circ}$. All 378 models were converted into additive tables using
the HEASoft task {\it ftflx2tab}. These were then used as one-parameter model components in XSPEC. All these components are referred to as reflection models in the following. We note here that the radius of the reflecting component is intended to simulate the presence of the accretion disk and the densest material in its vicinity \citep[e.g.,][]{Lee2022}. 
For instance, a similar analysis of T~CrB suggests that the best reflecting disk structure in this symbiotic recurrent nova has a radius of $R$=1.0 AU \citep{Toala_T_CrB_2024}, which is larger by a factor of $\approx 2$ than the accretion disk reported by \citet{Zamanov2024}.

An automatic procedure using Python routines was performed in XSPEC, adopting models of the form
\begin{equation}
{\rm tbabs}_1 \cdot ({\rm apec}_1 + {\rm apec_2}) + {\rm tbabs_2} \cdot {\rm CF} \cdot ({\rm apec_3} + {\rm reflection}),
\end{equation}
\noindent where CF is a covering factor that mimicks the nonuniform distribution of material around the accreting WD. 
The best models were assessed by comparing them with the observations and calculating the reduced $\chi^2$ statistics ($\chi^2/\mathrm{DoF}$).

As a first attempt, we performed a conjoint fit to all epochs by allowing all parameters to vary freely. However, this approach did not result in a consistent single disk model able to produce good quality fits ($\chi^2/\mathrm{DoF}<2$). The most problematic epochs correspond to 2013.96 and 2014.25. 
Thus, we started by producing a single joint fit to the 2014.84, 2015.74, 2015.75, and 2015.81 epochs, which resulted in a best-fit model ($\chi^{2}/\mathrm{DoF}=170.66/161=1.06$) for a disk model with $N_\mathrm{H,ref}=5\times10^{24}$~cm$^{2}$, $R$=1.25 AU, and a viewing angle $\theta=50^{\circ}$. 
Using this reflecting disk model, we performed individual fits to the 2013.96 and 2014.25 epochs, which also resulted in acceptable fit qualities ($\chi^{2}/\mathrm{DoF}$ equal to 1.67 and 1.25, respectively.

\begin{figure}
\centering
\includegraphics[width=1.0\linewidth]{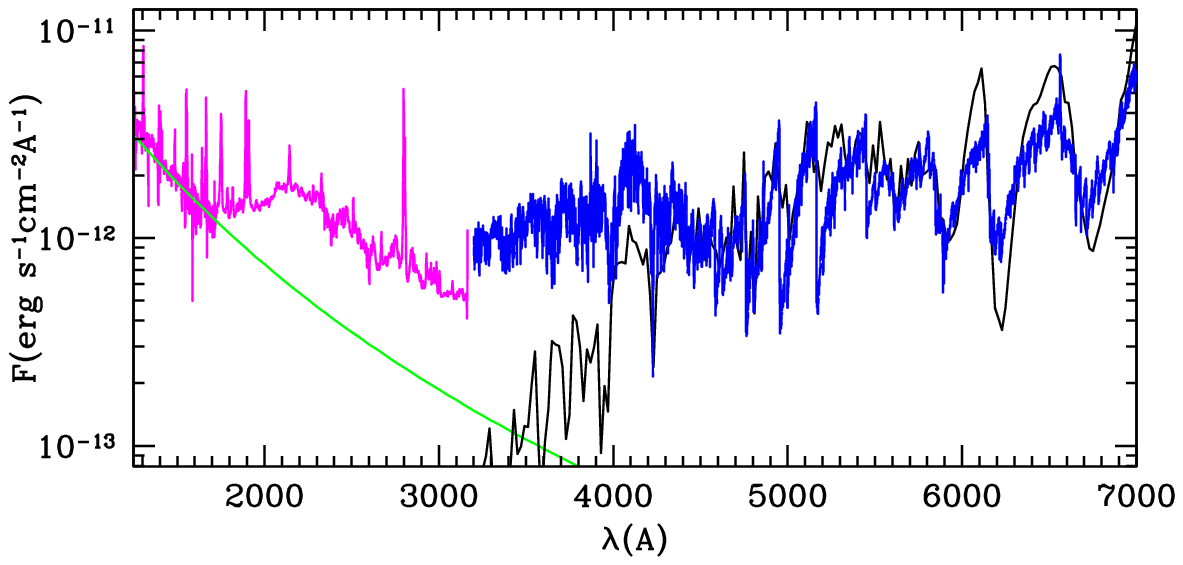}
\caption{UV and optical spectra of Y Gem. The figure includes far- and near-UV spectra from HST (magenta) and the optical INT IDS spectrum (blue), both corrected for extinction assuming $A_\mathrm{V}=0.8$ and the extinction curve by \citet{Cardelli+1989}. The black curve represents the best-fit spectrum from the library of stellar spectra by \citet{Lejeune+1997} corresponding to $T_{\rm eff}=3350$~K, $\log g = -0.29$, and [Fe/H]=0.0. The green curve represents a blackbody distribution with $T_{\rm eff}=60,000$ K and $L$=140~L$_{\odot}$ ($R = 0.11 R_{\odot}$).}
\label{fig:complete_spec} 
\end{figure}


The details of the best-fit models for all X-ray epochs are listed in Table~\ref{tab:analysis}. This table lists the total observed ($f_\mathrm{X,TOT}$) and intrinsic ($F_\mathrm{X,TOT}$) fluxes corresponding to the 0.3--10.0 keV energy range, as well as the contribution from each component. The total luminosities ($L_\mathrm{TOT,X}$) were computed adopting a distance of $d$=644 pc.

As expected, the 2013.96 epoch did not require any soft component to produce a good model. 
Similarly, the 2014.25 epoch only required one single thermal component to fit the soft X-ray range. The best-fits models of all epochs are also compared with the background-subtracted spectra in Fig.~\ref{fig:xspec}, and their details are further illustrated in Fig.~\ref{fig:xspec2}.

It is important to remark here that in some epochs the estimated flux of the reflection component ($F_\mathrm{ref}$) is on the same order of magnitude as that of the plasma temperature expected from the boundary layer between the accretion disk and the surface of the WD ($F_\mathrm{X3}$). The latter confirms the indisputably important role of the reflection component in X-ray-emitting SySts.

\begin{figure}
\centering
\includegraphics[width=1.0\linewidth]{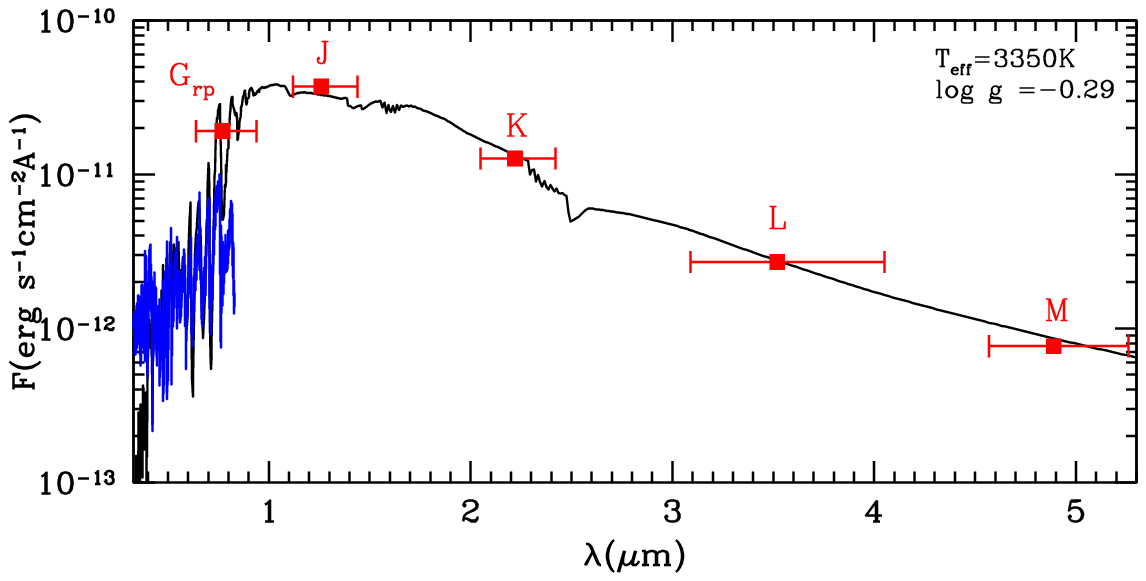}
\caption{INT IDS visual spectrum (blue) and near-IR flux densities (red squares) calculated from the {\it GAIA} and {\it DIRBE} multi-epoch photometry. The horizontal bars mark the FWHM of the broadband filters. All data were corrected for extinction assuming $A_\mathrm{V}=0.8$ and the extinction law by \citet{Cardelli+1989}. The black curve represents the best-fit synthetic spectrum from the library of \citet{Lejeune+1997}, and the parameters of the model are shown in the top right corner of the plot.}
\label{fig:SED} 
\end{figure}

\section{Discussion}
\label{sec:diss}

\subsection{The stellar components of Y Gem}

The optical spectrum of Y\,Gem exhibits the features of a cool star, but it also shows forbidden and H~{\sc i} recombination emission lines.  
The forbidden emission lines of [Ne~{\sc iii}] and [O~{\sc iii}] in particular suggest the presence of a hot companion, with $T_{\rm eff} \geq 30,000$~K. 
The optical spectrum of Y\,Gem is notably similar to that of the iconic SySts R\,Aqr, T\,CrB, and $o$~Cet. 
Thus, although some features typical of SySts are lacking in the optical spectrum of Y\,Gem, such as the He~{\sc ii} and O~{\sc vi} Raman-scattering emission lines, a SySt nature is certainly supported. 
To study the properties of the hot and cool components of Y\,Gem, we need to separate their possible contributions to its spectrum.

The multiwavelength spectra of Y\,Gem are presented in Fig.~\ref{fig:complete_spec}, including the UV spectrum obtained with HST STIS (1100--3200~\AA) and the optical INT IDS spectrum\footnote{It must be noted that Y\,Gem is highly variable, in particular, in the UV domain \citep{Sahai2011}. Its multiwavelength spectrum in Fig.~\ref{fig:complete_spec} may therefore be not representative of its average emission.}. 
The spectra were corrected for extinction using Cardelli's extinction law\footnote{Cardelli's extinction law is only valid above 1250 \AA, thus the bluest region of the HST STIS G140L spectrum was not considered for the subsequent analyses.} \citep{Cardelli+1989}, adopting a color excess $E$(B$-$V)=0.26 mag (or $A_\mathrm{V}$=0.80 mag), which corresponds to the extinction value derived from the Balmer line ratio in the optical INT IDS spectrum.

Fig.~\ref{fig:complete_spec} confirms that the red spectral region of Y\,Gem mostly presents emission from a cool, late-type component. Its spectral type can be estimated from its observed $(V-K)$ color of 8.0 mag \citep{Hog+2000,Cutri+2003}.
A total extinction (i.e., circumstellar plus interstellar) between $0.15 < A_V < 1.0$ implies a dereddened $7.1 < (V-K)_0 < 7.8$, which corresponds to $3000 < T_{\rm eff}$(K)$ < 3500$  and spectral type M5$-$7~{\sc ii}, depending on the spectrophotometric calibration adopted \citep{Ridgway1980,DiBenedetto1993,Perrin+1998,Pickles1998}, even though the stellar variability can widen the intervals above.

To refine the spectral classification of the cool component of Y\,Gem, its INT IDS optical spectrum in the range from 6300~\AA\ to 8300~\AA, which is expected to be dominated by the cool component, was subsequently compared with the library of synthetic spectra provided by \citet{Lejeune+1997} for a wide range of stellar parameters based on the atmosphere models by \citet{Bessell+1989}, \citet{Fluks+1994}, and \citet{AH1995}. 
Available broadband photometry of Y\,Gem provided by the multi-epoch monitoring DIRBE ($JKLM$) and Gaia (the $G_\mathrm{rp}$ band centered at 7830 \AA) was also used in this comparison.  
Following the results implied from its intrinsic $(V-K)_0$ color, the search focused on models of late-type giant stars ($\log g < +0.50$, 3000~K $< T_{\rm eff} <$ 3500~K) and solar metallicity.

The grid of stellar models of \citet{Lejeune+1997} provides the flux density $H_{\lambda}$ (in units of erg~s$^{-1}$~cm$^{-2}$~\AA~$^{-1}$) at the stellar surface for values of the stellar temperature and gravity spaced in intervals of 150--200~K and $\approx 0.3$~dex, respectively. 
Like in \citet{OG2016}, $H_{\lambda}$ was first convolved with the $G_\mathrm{rp}JKLM$ filter bandpasses $S(\lambda)$ \citep{Hauser1998,Jordi2010}  to obtain the surface flux relative to the $ith^{\rm }$ filter,

\begin{equation}
f_i = \frac{\int_{0}^{\infty} H(\lambda) \; S_X^i(\lambda) \;d \lambda}
{\int_{0}^{\infty} S_X^i(\lambda) \; d\lambda}.
\label{Fx}
\end{equation}

\noindent These flux densities were finally equally scaled (i.e. multiplied by a constant) in order to match the five extinction-corrected observed flux densities $F_i$. The best-fit model corresponds to the smallest total residual error, which was calculated as the sum of the squared individual differences of the five bands.
We found that two models show similar total residuals, namely $T_{\rm eff} = 3350$~K, $\log g = -0.51$ and $T_{\rm eff} = 3350$ K, $\log g = -0.29$, but the former solution can be discarded because it would imply a primary mass below $0.8 M_{\odot}$, which is unlikely for a star at this evolutionary stage. 
At any rate, the uncertainty in the stellar gravity (and consequently, in the stellar mass) derived by the method is significant because of the weak dependence on the stellar spectrum shape of stars such as Y\,Gem.
The synthetic spectrum is compared in Figs.~\ref{fig:complete_spec} and \ref{fig:SED} with the optical INT IDS spectrum and the available broadband $G_\mathrm{rp}JKLM$ extinction-corrected photometry, respectively.
The best-fit temperature and gravity uncertainties are at least as high as the pace between adjacent models, that is, $150-200$ K and 0.3 dex, respectively.

The integration of the best-fit scaled synthetic spectra over its full range (i.e., between 91 \AA \, and 160 $\mu$m) results in a flux of $5.15 \times 10^{-7}$ erg~cm$^{-2}$~s$^{-1}$. 
At the distance of 644 pc, this corresponds to a luminosity of $2.5 \times 10^{37}$ erg s$^{-1}$ or $6.6 \times 10^3 L_{\odot}$. For comparison, the luminosity computed from its $(J-K)$ color index \citep{Whitelock2000} would be $8.1 \times 10^3 L_{\odot}$, that is, a difference of $\approx 20\%$ from the luminosity derived from its best-fit spectrum, but within the expected uncertainty, considering its variability. The thermal pulse evolutionary tracks calculated by \citet{Marigo1996} in the H-R diagram for these luminosities and $T_{\rm eff}=3350$ K imply a main-sequence progenitor of $\approx$3 M$_{\odot}$. 
The present-day stellar mass is lower than this value because AGB stars lose a considerable part of their mass during the thermal-pulse phase. 
For example, theoretical models of late stellar evolution by \citet{MB2016} showed that a 3~M$_{\odot}$ main-sequence star with solar metallicity reaches the post-AGB phase with a final mass of $M_\mathrm{f} = 0.66$ M$_{\odot}$.

As long as the present-day mass of Y\,Gem can be roughly estimated from its stellar gravity and radius ($R$), the latter can be determined from its stellar luminosity ($L$) and temperature ($T_{\rm eff}$),

\begin{equation}
\log \left(\frac{R}{R_{\odot}}\right) = \frac{1}{2} \log \left(\frac{L}{L_{\odot}}\right) -
2 \log \left( \frac{T_{\rm eff}}{T_{\odot}}\right).
\end{equation}

\noindent Assuming $T_{\rm eff} = 3350$ K and $L/$L$_{\odot}= 6.6 \times 10^3$, we obtained $R/$R$_{\odot}= 240$, and the mass corresponding to this radius and $\log g = -0.29$ is 1.1 $M_{\odot}$. These quantities do not depend strongly on the extinction because an increment of 0.1 mag in $A_\mathrm{V}$ would increase the stellar luminosity and radius by 4\% and 1\%, respectively.

Theoretical models establish a relation between these stellar parameters and the pulsation mode of Mira- and SR-type variables \citep{OstlieCox1986,Wood1990}. The predicted radius of an AGB star that pulsates with period $P_0$ in the fundamental mode is

\begin{equation}
\log \left(\frac{R}{R_{\odot}}\right) = 0.54 \log P_0 + 0.39 \log \left(\frac{M}{M_{\odot}}\right) + 1.03,
\end{equation}

\noindent whereas if the star pulsates in the first-overtone mode with a period $P_1$, it is

\begin{equation}
\log \left(\frac{R}{R_{\odot}}\right) = 0.63 \log P_1 + 0.32 \log \left(\frac{M}{M_{\odot}}\right) + 1.01.
\end{equation}

\noindent
For a stellar mass of 1.1 M$_{\odot}$, the stellar radius would be 165 and 245 R$_{\odot}$ for the fundamental and first-overtone mode. Therefore, the stellar parameters of Y\,Gem obtained above indicates that it pulsates in the first overtone, like the SR variables in the LMC \citep{Wood1999}.

The extinction-corrected HST STIS UV spectrum in Fig.~\ref{fig:complete_spec} shows a notorious blueward ramp starting at $\approx$1700 \AA, which is suggestive of a hot stellar component. Fitting the slope of this far-UV spectrum with a blackbody requires a minimum temperature of $\approx 60,000$ K, which agrees with the detection of 
the [O~{\sc iii}] and [Ne~{\sc iii}] emission lines.  
Otherwise, its $T_{\rm eff}$ cannot significantly exceed a value $\simeq$54,000~K, as the He~{\sc ii} $\lambda$4686 emission line is undetected in the optical spectrum. 
Therefore, a hot component with $T_{\rm eff}\approx 60,000$ K seems to be a good compromise between the far-UV spectral slope and the atomic species in the optical spectrum. 
The radius of a hot stellar component like this that would fit the far-UV spectrum is 0.11 R$_{\odot}$, which corresponds to 140 L$_{\odot}$. We note that accreting WD are known to have much larger radii than the typical noninteracting WDs \citep[e.g.,][]{Livio1989,Burgarella1992}.

The nondetection of the He~{\sc ii} $\lambda$4686 emission line seems to indicate an effective temperature of the hot companion that is lower than 54,000 K.  
This requires a shallower far-UV ramp, which can be reconciled with the observed one if the extinction of $A_V$ equal to 0.8 mag used for its reddening correction were lower.  
This possibility cannot be neglected, since self-absorption of the Balmer lines, as discussed in Sect.~3.1, can mimic a higher extinction.  
For comparison, the HST STIS UV spectrum was fit by \citet{Sahai+2018} without any reddening correction using two blackbody-emission models with $T_\mathrm{eff,1}$=35,000~K and $T_\mathrm{eff,2}$=9400 K, respectively. 
The effective temperature of the hot companion is thus constrained to be in the range of 35,000~K to 54,000~K.

It it worth mentioning that \citet{Sahai+2018} concluded that the two stellar components required by their spectral fit should  originate in an accretion disk. 
Moreover, they attributed blue absorption of the Ly$\alpha$, N~{\sc v}, O~{\sc i}, Si~{\sc iv}, and C~{\sc iv} lines to an outflow and assigned the observed radial velocity to an escape velocity to suggest that the accreting star is a low-mass main-sequence star.
The forbidden emission lines in the optical spectrum of Y\,Gem and its X-ray emission, however, support a hot accreting WD companion. 
Otherwise, our best-fit model of the reflecting properties of Y\,Gem implies an inclination of 50$^{\circ}$ with respect to the orbit of the binary system. 
The true velocity of the outflow can be as high as $\approx$1500~km~s$^{-1}$ when the observed radial velocity $\approx-1000$ km~s$^{-1}$ of the Ly$\alpha$ line in the UV spectra of Y Gem \citep{Sahai+2018} is corrected for a tilt of the outflow of $50^\circ$ with the line of sight. 
Using the definition of the escape velocity,
\begin{equation}
\varv_\mathrm{esc} = \sqrt{\frac{2 G M}{R}},
\end{equation}
\noindent 
we found that a WD with a mass of 0.8~M$_\odot$ \citep[as estimated by][]{Yu+2022} and $R$=0.11~R$_\odot$ results in a very similar value of $\approx$1670~km~s$^{-1}$ for the escape velocity.

\subsection{Mass-accretion rate}

We can assess the accretion disk luminosity by subtracting the contributions of the hot WD companion and the late M-type star from the UV and optical spectrum of Y\,Gem. 
This emission excess is estimated to be $L_\mathrm{disk}$=28~L$_{\odot}$.

Assuming that the accretion process produces a total accretion luminosity ($L_\mathrm{acc}$), 
we can approximate the mass-accretion rate $\dot{M}_\mathrm{acc}$ as \citep[see][]{Shakura1973,Pringle1981}
\begin{equation}
L_\mathrm{acc} = \frac{G M_\mathrm{WD}}{R_\mathrm{WD}} \frac{\dot{M}_\mathrm{acc}}{2},     
\label{eq:acc}
\end{equation}
\noindent with $R_\mathrm{WD}$ as the radius of the WD component and, in our case, the best approximation for the accretion luminosity, that is, $L_\mathrm{acc} \approx L_\mathrm{disk} + L_\mathrm{X}$.

At this point, it is paramount to account for the appropriate X-ray luminosity. 
Our detailed modeling of the X-ray observations of Y\,Gem found different components that contribute to its X-ray spectra. 
As found in other $\beta/\delta$-type X-ray-emitting SySts \citep[e.g.,][]{Lucy2020,Toala2022}, the soft X-ray emission (spectral fit components $kT_1$ and $kT_2$) of Y\,Gem very likely arises from adiabatically shocked regions due to the high velocity and variable outflows in Y\,Gem \citep{Sahai+2018}. 
In addition, the reflection component detected in all epochs is not produced by the accretion process either. 
Therefore, only the heavily extinguished plasma component, $kT_3$, typically associated with the plasma temperature of the boundary layer between the accretion disk and the surface of the WD, should be considered to compute the accretion luminosity.

During the different X-ray observation epochs, the luminosity of the $kT_3$ plasma component ($L_\mathrm{X3}$) ranged from 0.05 to 0.15~L$_{\odot}$, thus, in order to estimate the mass-accretion rate, we adopted an averaged value of 0.12~L$_{\odot}$ for the X-ray contribution. 
Using then Eq.~(\ref{eq:acc}), we estimate $\dot{M}_\mathrm{acc}$= 2.5$\times10^{-7}$~M$_{\odot}$~yr$^{-1}$ adopting the radius of 0.11~R$_\odot$ estimated for the accreting WD in Y Gem and a mass of 0.8~M$_\odot$.
Based on the comparison of our $\dot{M}_\mathrm{acc}$ estimate with theoretical predictions for accreting WDs, the WD component in Y\,Gem appears to have reached the stable and steady burning phase described for a 0.8 M$_\odot$ \citep{Cassisi1998,Wolf2013}, which seems to prevent a recurrent phase for this system.

\subsection{Consequences of Y Gem as an SySt}

The results discussed in the previous sections confirm the status of Y\,Gem as an SySt \citep{Yu+2022} and not an AGB star with an accreting main-sequence companion \citep[see][and references therein]{Sahai+2018}. 
We further used the available $JHK$ and WISE mid-IR magnitudes to explore the classification trees and color-color diagrams presented by \citet{Akras+2019}.  
These confirm that Y\,Gem is consistent with an S-type SySt.

The properties of the accreting WD companion, however, are only poorly constrained.  
The recombination and forbidden emission lines detected in the optical spectrum constrain $T_\mathrm{eff}$ between 30,000 and 60,000~K, but the UV spectrum blueward of 1700~\AA\ requires $T_\mathrm{eff}$ to be $\geq$60,000~K. 
We note that in some cases, the optical spectra of SySts are extremely variable, with bright recombination lines in their spectra that completely disappear at different epochs (see, e.g., the cases of NQ\,Gem and RT\,Cru in the ARAS spectral database).
This might be the case of Y\,Gem as well, given its extraordinary UV and X-ray variability.

The multiwavelength analysis of Y\,Gem presented here challenges our understanding of X-AGBs. 
It poses the question whether all X-AGB systems are unidentified SySts. 
\citet{Guerrero+2024} recently discussed these two populations by comparing their UV and X-ray properties. 
Both X-AGBs and SySts seem to share similar X-ray luminosity values, with confirmed SySts having a typical value of
$\log(L_\mathrm{X}/\mathrm{erg}~\mathrm{s}^{-1})$=32.0$\pm$0.8, while for the distribution of so-called X-AGBs, they found $\log(L_\mathrm{X}/\mathrm{erg}~\mathrm{s}^{-1})$=30.6$\pm$0.8. 
Regardless of the evolution of the X-ray properties of Y\,Gem that it exhibited during the XMM-Newton and Chandra observations analyzed here, its total X-ray luminosity is definitely more consistent with the upper limit estimated for SySts.  
We used the HST STIS UV spectra of Y\,Gem to estimate its far-UV magnitude, for which we adopted the GALEX transmission curve. We found it to be M$_\mathrm{[FUV]}$=1.63 mag. 
This means that Y\,Gem is one of the brightest UV and X-ray SySts \citep[see the compilation presented in figure 4 of][]{Guerrero+2024}.

\citet{Guerrero+2024} suggested that the lower X-ray luminosity range defined for X-AGBs seems to be powered by accreting main-sequence stars, but this statement is challenged by $o$~Cet, a SySt in the locus of X-AGBs in the M$_\mathrm{[FUV]}$ versus $L_\mathrm{X}$ diagram.  
The low X-ray luminosity of $o$~Cet has been attributed to a low-accretion rate onto the WD resulting from a wide orbit with period $\approx$500 yr\footnote{see the details of $o$ Cet in the New Online Database of Symbiotic Variables \url{https://sirrah.troja.mff.cuni.cz/~merc/nodsv/stars/omi_cet.html}}, which might be the case of other X-ray weak SySts. 
We expect that future multiwavelength studies of X-AGB stars will help us to unveil unidentified X-ray-emitting SySts in the low-luminosity range of the X-ray regime.

Finally, the orbital parameters of the binary system formed by the late 1.1~M$_\odot$ Y\,Gem and its 0.8 M$_\odot$ WD companion can be assessed. 
Adopting the shortest possible period for the binary obtained from the AAVSO light-curve analysis ($P_\mathrm{WD1?}$=8.87 yr), we find that the semimajor axis of the system results in $a \approx 5.3$ AU for an effective Roche-lobe radius of 2.2 AU.  
The Roche lobe is thus larger than the stellar radius of the primary component of 240 R$_\odot$(=1.12 AU), that is, the primary cannot fill its Roche lobe. 
This suggests that the accretion mechanism in Y\,Gem works through a Bondi-Hoyle-Lyttleton (BHL) scenario \citep{Hoyle1939,Bondi1944,Bondi1952} or through the so-called wind Roche-lobe overflow channel \citep[e.g.,][]{Podsiadlowski2007}.
This is also the case for the other longer periods ($P_\mathrm{WD2?}$ and $P_\mathrm{WD3?}$), which imply larger orbital separations (11.7 and 20.3 AU).

Adopting $P_\mathrm{WD1?}=8.87$ yr as the binary period, we used the modify BHL model recently presented by \citet{TejedaToala2024} to assess the mass-accretion rate of the WD component in Y\,Gem. 
For the most probable stellar wind velocity of 5~km~s$^{-1}$ and typical mass-loss rates of $\dot{M}$=[0.7--1]$\times10^{-6}$~M$_\odot$~yr$^{-1}$ of SR-type stars \citep[e.g.,][]{Olofsson2002}, we found a mass-accretion rate of $\dot{M}_\mathrm{acc}\gtrsim10^{-7}$. 
This value is consistent with the value estimated from the X-ray observations, and it favors the shorter orbital $P_\mathrm{WD1?}$ period over the longer $P_\mathrm{WD2?}$ and $P_\mathrm{WD3?}$ periods of 28.9 and 65.9 yr, respectively, which can be expected to result in lower mass transfer efficiencies from the late-type star wind onto the WD component.

\section{Conclusions}
\label{sec:conclusions}

We presented the analysis of optical, UV, and X-ray observations of Y\,Gem to confirm that this previously classified X-AGB star is in fact an SySt \citep[as suggested by][]{Yu+2022}. 
Our optical spectroscopic observations showed that Y\,Gem shares very similar spectral properties with other iconic SySts such as $o$ Cet, R~Aqr, and T CrB. 
The near- and mid-IR colors indicate that Y\,Gem is an S-type SySt.

The recombination and forbidden lines in the optical spectrum require a temperature of the companion star $\geq$30,000~K, which is impossible for a main-sequence star companion of an AGB, but typical of a WD.  
The far-UV spectrum requires an even higher temperature $\geq$60,000~K for the hot component. 
The analysis of the optical spectrum and available IR photometry helped us to constrain the properties of the late-type companion, which is a giant AGB star with an effective temperature of 3350~K, a radius of 240 R$_\odot$, and a mass of 1.1 M$_\odot$.
After subtracting the contribution from the hot and cool stellar components, we estimate a UV+optical luminosity for the accretion disk of $L_\mathrm{disk}=$28~L$_\odot$.

Publicly available X-ray observations were interpreted by means of a reflection component. 
In this scenario, X-ray photons produced at the boundary layer between the accretion disk and the surface of the WD are scattered and reflected by material in the accretion disk and by material in the vicinity of the SySt. 
This analysis allowed us to separate the different contributions to the X-ray emission detected from Y\,Gem. 
Independent estimates for the soft component (most likely caused by shocks associated with variable jet-like ejections; $L_\mathrm{soft}=L_\mathrm{X1} + L_\mathrm{X2}$), the reflecting component ($L_\mathrm{ref}$), and the actual X-ray emission produced at the boundary layer as a consequence of accretion ($L_\mathrm{X3}$).
Consequently, we estimate that the current accretion luminosity can be approximated by $L_\mathrm{acc}=L_\mathrm{disk} + L_\mathrm{X3}$, which can be used to estimate a mass-accretion rate $\dot{M}_\mathrm{acc}=2.5\times$10$^{-7}$~M$_\odot$~yr$^{-1}$. 
By comparing this with theoretical models of accreting WDs, its seems that Y\,Gem has reached a stable and steady burning phase where no recurrent events are expected.

Our study revealed that Y\,Gem is one of the brightest UV and X-ray SySts reported so far.  
Its conversion from an AGB star into an X-AGB star and finally to an SySt encourages more searches and multiwavelength characterizations of X-AGB stars, which might unveil the long-sought missing population of SySts \citep[see][]{Petit2023,Xu+2024}.

\begin{acknowledgements}
The authors thank the anonymous referee for comments and suggestions that improved the analysis and presentation of our results, and François Teyssier for sharing with us the ARAS spectrum of Y\,Gem whose prompt observation followed the publication of a preliminary version of this work on arXiv.  
M.A.G.\ acknowledges financial support from grants CEX2021-001131-S funded by MCIN/AEI/10.13039/501100011033 and PID2022-142925NB-I00 from the Spanish Ministerio de Ciencia, Innovaci\'{o}n y Universidades (MCIU) cofunded with FEDER funds.
J.B.R.G.\ and D.A.V.T.\ thank CONAHCyT (Mexico) for student grants. 
J.B.R.G., D.A.V.T., and J.A.T.\ are supported by Universidad Nacional Aut\'{o}noma de M\'{e}xico (UNAM) PAPIIT project IN102324. 
J.A.T.\ also thanks the Centro de Excelencia Severo Ochoa Visiting-Incoming programme for support during a visit to IAA-CSIC (Spain). 
R.O.\ thanks the support of the S\~ao Paulo Research Foundation (FAPESP), grant \#2023/05298-0.
This work is based on service observations made with the Isaac Newton Telescope (programme SST2024-654) operated on the island of La Palma by the Isaac Newton Group of Telescopes in the Spanish Observatorio del Roque de los Muchachos of the Instituto de Astrofísica de Canarias. 
The observer, R.\ Clavero, telescope operator, E.\ Mantero, and support astronomer, Santos, are particularly acknowledged.  
This work is based on observations obtained with XMM–Newton, an European Science Agency (ESA) science mission with instruments and contributions directly funded by ESA Member States and NASA. The scientific results reported in this paper are based on observations made by the Chandra X-ray Observatory and published previously 
in cited articles. 
This research is based on observations made with the NASA/ESA Hubble Space Telescope obtained from the Space Telescope Science Institute, which is operated by the Association of Universities for Research in Astronomy, Inc., under NASA contract NAS 5–26555. These observations are associated with program 14713. We acknowledge with thanks the variable star observations from the AAVSO International Database contributed by observers worldwide and used in this research.
This work has made extensive use of NASA's Astrophysics Data System. 
\end{acknowledgements}

%
%


\end{document}